\documentclass[12pt]{article}
\usepackage{epsfig,exscale}
\IfFileExists{hyperref.sty}{\usepackage{hyperref}}{}

\thicklines

\newtheorem{Lemma}{Lemma}
\newtheorem{Proposition}[Lemma]{Proposition}

\def\qed{\leavevmode \hfill\hbox to.77778em{\hfil\vrule
         \vbox to.675em{\hrule width.6em\vfil\hrule}\vrule\hfil}}

\def\semi{\mathop{>\!\!\!\triangleleft}}

\makeatletter
\def\section{\@startsection{section}{1}{\z@}{-3.25ex 
       plus -1ex minus -.2ex}{1.5ex plus .2ex}{\normalfont\bfseries}}
\def\subsection{\@startsection{subsection}{2}{\z@}{-3.25ex 
        plus -1ex minus -.2ex}{1.5ex plus .2ex}{\normalfont\itshape}} 

\def\thebibliography#1{\section*{\refname}\list
  {\@biblabel{\theenumiv}}{\settowidth\labelwidth{\@biblabel{#1}}%
    \leftmargin\labelwidth
    \advance\leftmargin\labelsep
    \footnotesize \parsep=0pt \itemsep=0pt
    \usecounter{enumiv}%
    \let\p@enumiv\@empty
    \def\theenumiv{\arabic{enumiv}}}%
    \def\newblock{\hskip .11em plus.33em minus.07em}%
    \sloppy\clubpenalty4000\widowpenalty4000
    \sfcode`\.=1000\relax}

\def\ps@hepth{\addtolength{\headheight}{5pt}
            \addtolength{\topmargin}{-15pt}
            \addtolength{\headsep}{15pt}
    \def\@oddhead{\hfil\begin{tabular}{r}
          \texttt{hep-th/9912221} \\ 
          CPT-99/P.3814
          \end{tabular}}
          \let\@evenhead\@oddhead
          \def\@oddfoot{\hfil\thepage\hfil}\let\@evenfoot\@oddfoot}

\makeatother

\font\twelvebb=msbm10 scaled 1200
\font\eightbb=msbm8 scaled 1000
\newfam\bbfam
\textfont\bbfam=\twelvebb
\scriptfont\bbfam=\eightbb
\def\bb{\fam\bbfam}

\begin{document}

\thispagestyle{hepth}

\begin{center}

{\baselineskip 20pt 
{\large{\bfseries Introduction to Hopf algebras in
renormalization and noncommutative geometry} 
\footnote{based on talk given at the Hesselberg'99 conference 
March 15--19, 1999}}
\vskip 3ex

{\sc Raimar Wulkenhaar} \footnote{e-mail
\texttt{raimar@cpt.univ-mrs.fr}}\footnote{supported by the German
Academic Exchange Service (DAAD), grant no.\ D/97/20386} 
}
\vskip 1ex 
 
{\itshape Centre de Physique Th\'eorique\\
CNRS-Luminy, Case 907\\
13288 Marseille Cedex 9, France}
\end{center}
\vskip 3ex

\begin{abstract}
We review the appearance of Hopf algebras in the renormalization of
quantum field theories and in the study of diffeomorphisms of the
frame bundle important for index computations in noncommutative
geometry.
\end{abstract}

\vskip 1ex

\section{Introductory remarks}

This contribution focuses on two applications discovered during the
last two years of the Hopf algebra of rooted trees. They suggest an
amazing link between mathematics and physics. There exists an
excellent review \cite{ck2} of these topics, written by the authors of
these ideas. In what follows I am going to explain parts of this
development which I was able to understand. I hope it can be useful to
somebody else.

In mathematics, foliations provide a large class of examples of
noncommutative spaces and lead to an index problem for the transverse
hypo\-elliptic operator \cite{cm1}. The computation of the cocycles in
the local index formula turned out to be extremely lengthy even in
dimension one. Alain Connes and Henri Moscovici \cite{cm2} were
looking for an organizing principle for that calculation, which they
found in the cyclic cohomology of a Hopf algebra $\mathcal{H}\sb T$
obtained by the action of vector fields on a crossed product of
functions by diffeomorphisms.

Concerning physics, Dirk Kreimer \cite{k1} discovered that a
perturbative quantum field theory carries in a natural way a Hopf
algebra structure $\mathcal{H}\sb R$ given by operations on Feynman
graphs. The antipode reproduces precisely the combinatorics of
renormalization, i.e.\ it produces the local counterterms to make the
divergent integral corresponding to the Feynman graph finite.

Noticing that both Hopf algebras have formally a very similar
structure, Connes and Kreimer gave the precise relation \cite{ck1}
between $\mathcal{H}_T$ and $\mathcal{H}_R$. This is very transparent
in the language of rooted trees they used. The commutative Hopf
subalgebra $\mathcal{H}^1$ of Connes--Moscovici is (in dimension 1) a
Hopf subalgebra of Kreimer's Hopf algebra for a quantum field theory
with a single primitively divergent graph.

Recently it was pointed out \cite{b} that the same algebra of rooted
trees plays a role in Runge--Kutta methods of numerical analysis.

\section{The Hopf algebra of Connes--Moscovici}

In principle, the Hopf algebra of Connes and Moscovici can be
understood from classical differential geometry \cite{cd}. We give
here a somewhat shortened version of the derivation and refer to
\cite{rw} for more details. We recommend \cite{m} for a useful
introduction to Hopf algebras and related topics.

We regard the frame bundle $F^+$ of a manifold $M$ and in particular
the vector fields on $F^+$. There is a natural notion of vertical
vector fields, these are the tangent vectors to curves in $F^+$
obtained by the right action of the group $Gl^+(n)$ of $n\times n$
matrices with positive determinant. The horizontal vector fields are
not canonically given, they are determined once a connection is
specified. For our purpose we can work in local coordinates.

Let $\{x^\mu\}_{\mu=1\dots,n}$ be the coordinates of $x\in M$ within a
local chart of $M$ and $\{y^\mu_i\}_{\mu,i = 1,\dots n}$ be the
coordinates of $n$ linearly independent vectors of the tangent space
$T_xM$ with respect to the basis $\partial_\mu$. On $F^+$ there exist
the following geometrical objects, written in terms of the 
local coordinates $(x^\mu,y^\mu_i)$ of $p \in F^+$:
\begin{itemize}
\item[1)] an ${\bb R}^n$-valued (soldering) $1$-form $\alpha$ with
$\alpha^i = (y^{-1})^i_\mu d x^\mu~,$

\item[2)] a $gl(n)$-valued  (connection) $1$-form $\omega$ with 
$\omega^i_j = (y^{-1})^i_\mu ( d y^\mu_j + \Gamma^\mu_{\alpha\beta}\,
y^\alpha_j dx^\beta)$, where $\Gamma^\mu_{\alpha\beta}$ depends only
on $x^\nu~,$

\item[3)] $n^2$ vertical vector fields $Y^i_j = y^\mu_j
\partial^i_\mu~,$

\item[4)] $n$ horizontal (with respect to $\omega$) vector fields 
$X_i = y^\mu_i ( \partial_\mu - 
\Gamma^\nu_{\alpha\mu} y^\alpha_j \partial^j_\nu)~.$
\end{itemize}
A local diffeomorphism $\psi$ of $M$ has a lift $\tilde{\psi}:
(x^\mu,y^\mu_i) \mapsto (\psi(x)^\mu,\partial_\nu \psi(x)^\mu
y^\nu_i)$ to the frame bundle and induces the following
transformations of the previous geometrical objects:
\begin{itemize}
\item[$1')$] $(\tilde{\psi}^* \alpha)\big|_p = \alpha\big|_p~.$
\item[$2')$] $(\tilde{\psi}^* \omega)\big|_p = (y^{-1})^i_\mu ( d
y^\mu_j + \tilde{\Gamma}^\mu_{\alpha\beta}\, y^\alpha_j dx^\beta)$ is
again a connection form, with \\[0.5ex]
$\tilde{\Gamma}^\mu_{\alpha\beta}\big|_x =
((\partial\psi(x))^{-1})^\mu_\gamma \,
\Gamma^\gamma_{\delta\epsilon}\big|_{\psi(x)} \,
\partial_\alpha\psi(x)^\delta \partial_\beta \psi(x)^\epsilon +
((\partial\psi(x))^{-1})^\mu_\gamma \, \partial_\beta \partial_\alpha
\psi(x)^\gamma ~,$

\item[$3')$] $(\tilde{\psi}_* Y^j_i)\big|_p = Y^j_i\big|_p~,$

\item[$4')$] $(\tilde{\psi}^{-1}_* X_i)\big|_p = y^\mu_i 
( \partial_\mu - \tilde{\Gamma}^\nu_{\alpha\mu} y^\alpha_j 
\partial^j_\nu)$ is horizontal to $\tilde{\psi}^* \omega~.$
\end{itemize}
We refer to \cite{rw} for the proof.

Given these tools of classical differential geometry, the new idea is
to apply the vector fields $X,Y$ to a crossed product ${\cal
A}=C_c^\infty(F^+) \semi \Gamma$ of the algebra of smooth functions on
$F^+$ with compact support by the action of the pseudogroup $\Gamma$
of local diffeomorphisms of $M$. As a set, $\mathcal{A}$ can be regarded
as the tensor product of $C^\infty_c(F^+)$ with $\Gamma$. It is
generated by the monomials
\begin{equation}
f U^*_\psi~,\qquad f \in C^\infty_c({\rm Dom}(\tilde{\psi}))~,\quad 
\psi \in \Gamma~,
\end{equation}
where $\tilde{\psi}$ is the diffeomorphism of $F^+$ obtained as the
lift of $\psi \in \Gamma$. As an algebra, the multiplication rule in
$\mathcal{A}$ is defined by
\begin{equation}
f_1 U^*_{\psi_1} \; f_2 U^*_{\psi_2} := f_1 (f_2 \circ \tilde{\psi}_1) 
U^*_{\psi_2 \psi_1} ~.
\label{prod}
\end{equation}
Here, the function $f_1 (f_2 \circ \tilde{\psi}_1)$ evaluated at $p$
(in the domain of definition) gives $f_1(p) \,f_2
(\tilde{\psi}_1(p))$, i.e.\ we have a non-local product on the
function algebra.

The action of vector fields on $\mathcal{A}$ is defined as the action
on the function part. Interesting is the application to the product
(\ref{prod}), because the non-locality in the function part leads to a
deviation from the Leibniz rule. For $V$ being a vector field on $F^+$
one computes
\begin{eqnarray}
V(f_1 U^*_{\psi_1} \; f_2 U^*_{\psi_2}) =
V(f_1 U^*_{\psi_1}) \; f_2 U^*_{\psi_2} 
+ f_1 U^*_{\psi_1} \; \big(\tilde{\psi}_{1*}
(V) \big) \big(f_2 U^*_{\psi_2}\big) ~.
\label{Vab}
\end{eqnarray}
Since diffeomorphisms and right group action commute, we get the
unchanged Leibniz rule for the vertical vector fields,
\begin{equation}
Y^j_i(ab) = Y^j_i(a) \,b + a\, Y^j_i(b) ~,\qquad a,b \in \mathcal{A}~.
\label{Yab}
\end{equation}
For the horizontal vector fields, however, there will be an additional
term $a ({\psi_1}_* X_i-X_i)(b)$. Comparing $4)$, $4')$ and $3)$ above
we have ${\psi_1}_* X_i-X_i = \tilde{\delta}^k_{ji} Y^j_k$, for some
function $\tilde{\delta}^k_{ji}$. Using (\ref{prod}) we commute this
function in front of $a$ and obtain
\begin{equation}
X_i(ab) = X_i(a)\,b + a\,X_i(b) + \delta^k_{ji}(a)\,Y^j_k(b)~,\qquad
a,b \in \mathcal{A}~.
\label{Xiab}
\end{equation}
The operator $\delta^k_{ji}$ on $\mathcal{A}$ is computed to 
\begin{equation}
\delta^k_{ji} (f U^*_\psi) = (\tilde{\Gamma}^\nu_{\alpha\mu}  -
\Gamma^\nu_{\alpha\mu}) y^\alpha_j y^\mu_i (y^{-1})^k_\nu f U^*_\psi~,
\label{gamma}
\end{equation}
where $\tilde{\Gamma}^\nu_{\alpha\mu}$ are the connection coefficients
belonging to $\tilde{\psi}^* \omega$. It turns out that
$\delta^k_{ji}$ is a derivation:
\begin{equation}
\delta^k_{ji}(ab) = \delta^k_{ji}(a)\,b + a\,\delta^k_{ji}(b)~.
\label{delta}
\end{equation}

These formulae can now be interpreted in the dual sense, for instance
$X_i (ab) = \Delta(X_i)\,(a \otimes b)$, which leads to a structure of
a coalgebra on the linear space ${\bb R}(1,X_i,Y_k^j,\delta^k_{ji})$,
\begin{eqnarray}
\Delta(Y^j_k) &=& Y^k_j \otimes 1 + 1 \otimes Y^j_k~, \nonumber 
\\
\Delta(X_i) &=& X_i \otimes 1 + 1 \otimes X_i + \delta^k_{ji} \otimes 
Y^j_k~, \label{Delta}
\\
\Delta(\delta^k_{ji}) &=& \delta^k_{ji} \otimes 1 
+ 1 \otimes \delta^k_{ji}~, \nonumber 
\\
\Delta(1) &=& 1 \otimes 1~, \nonumber
\end{eqnarray}
with $1$ being the identity on $\mathcal{A}$. Coassociativity 
$(\Delta \otimes {\rm id}) \circ \Delta = ({\rm id} \otimes \Delta) 
\circ \Delta$ is easy to check.

Vector fields form a Lie algebra, so the next step is to ask whether
${\bb R}(1,X_i,Y_k^j,\delta^k_{ji})$ close under the Lie bracket. The
first commutators are OK,
\begin{eqnarray}
[Y^i_j,Y^k_l] (fU^*_\psi) &=& (\delta^i_l Y^k_j - \delta^k_j Y^i_l) 
(fU^*_\psi)~, \nonumber
\\ 
{}[Y^k_j,X_i] (fU^*_\psi) &=& \delta^k_i X_j (fU^*_\psi)~,
\\ {}
[Y^i_j,\delta^k_{lm}] (fU^*_\psi)  &=&
(\delta^i_l \delta^k_{jm} + \delta^i_m \delta^k_{lj} - 
\delta^k_j \delta^i_{lm}) ( fU^*_\psi)~. \nonumber
\end{eqnarray}
The next one between horizontal fields 
\begin{equation}
[X_i,X_j] = R^k_{lij} Y^l_k + \Theta^k_{ij} X_k
\end{equation}
leads to new generators, because curvature $R$ and torsion $\Theta$
are no structure `constants'. Therefore, one uses a different strategy
and considers instead of $\mathcal{A}$ a \emph{Morita equivalent}
algebra $\mathcal{A}'$ based on a \emph{flat} manifold $N=\coprod
U_\alpha$ -- the disjoint union of the charts $U_\alpha$ of $M$. Now,
there is neither curvature nor torsion, and horizontal vector fields
commute. There remain the commutators of $X$ with $\delta$, which lead
indeed to new generators of the Lie algebra:
\begin{eqnarray}
\delta^k_{ji,\ell_1\dots \ell_n} (f U^*_\psi) &:=& 
[X_{\ell_n},\dots ,[X_{\ell_1},\delta^k_{ji}]\dots] (f U^*_\psi) 
\label{flat}
\\ \nonumber 
&=& \partial_{\lambda_n} \dots \partial_{\lambda_1} 
\Big( \!((\partial\psi(x))^{-1})^\nu_\beta\,
\partial_\mu\partial_\alpha \psi(x)^\beta \Big) y^\mu_j y^\alpha_i
(y^{-1})^k_\nu\, y^{\lambda_1}_{\ell_1} \cdots y^{\lambda_n}_{\ell_n} 
\;  f U^*_\psi\,.
\end{eqnarray}
All these generators $\delta^k_{ji,\ell_1\dots \ell_n}$ commute with
each other.

Now having established a Lie algebra, we call $\mathcal{H}$ its
enveloping algebra, i.e.\ the algebra of polynomials in
$\{1,X_i,Y^k_j,\delta^k_{ji},\delta^k_{ji,\ell_1 \dots \ell_n
\dots}\}$, with the commutation relations inherited from the Lie
algebra. With the coproduct $\Delta$ on the Lie algebra, $\mathcal{H}$
becomes automatically a bialgebra, where the coproduct is defined via
the algebra homomorphism axiom:
\begin{equation}
\Delta (h^1 h^2) = \Delta (h^1) \, \Delta(h^2) := \sum h_1^1 h^1_2
\otimes h_1^2 h^2_2 ~,\qquad \Delta(h_i) = \sum h^1_i \otimes h^2_i~,
\label{copro}
\end{equation}
for $h_1,h_2 \in \mathcal{H}$. The counit $\epsilon: \mathcal{H} \to
{\bb C}$ is defined by
\begin{equation}
\varepsilon(1)=1_{\bb C}~,\qquad \varepsilon(h)=0\quad 
\forall h\neq 1 ~.
\end{equation}
The counit axiom $(\varepsilon \otimes {\rm id}) \circ \Delta(h) = 
({\rm id} \otimes \varepsilon) \circ \Delta(h) = h$ 
is straightforward to check.

There also exists an antipode on $\mathcal{H}$ which makes it to a
Hopf algebra. The antipode is the unique antiautomorphism of
$\mathcal{H}$ satisfying
\begin{eqnarray}
S(h_1 h_2) &=& S(h_2) S(h_1)~,
\nonumber \\
m \circ (S \otimes {\rm id}) \circ \Delta(h) &=& 1 \varepsilon(h) 
= m \circ ({\rm id} \otimes S) \circ \Delta(h) ~,\label{Sax}
\end{eqnarray}
for $h,h_1,h_2 \in \mathcal{H}$, and where $m$ denotes the
multiplication. From the second line and (\ref{Delta}) one easily
obtains
\begin{eqnarray}
S(1) &=& 1 ~, \nonumber \\
S(Y^j_k) &=& - Y^j_k~, \nonumber \\
S(\delta^k_{ji}) &=& - \delta^k_{ji}~, \\
S(X_i) &=& - X_i + \delta^k_{ji} Y^j_k~. \nonumber
\end{eqnarray}
The action of $S$ on the other generators of $\mathcal{H}$ can be
derived from (\ref{Sax}).

The purpose of this Hopf algebra $\mathcal{H}$ is to ease the
computation \cite{cm2} of cocycles in the local index formula
\cite{cm1} of Connes and Moscovici. So far I did not study this
calculation for myself, but I think a good way to learn it would be to
consult \cite{c}.

\section{Rooted trees}

Coproduct and antipode for the generators $\delta^k_{ji,\ell_1\dots
\ell_n \dots}$ are only recursively defined via the axioms of
coproduct and antipode. Now we are going to present an explicit
solution -- via the concept of rooted trees. This was introduced by
Connes and Kreimer \cite{ck1} to clarify the relation between the two
Hopf algebras in the theory of foliations and in perturbative quantum
field theory. We generalize \cite{rw} their construction from
dimension 1 to arbitrary dimension of the manifold $M$. To the first
three classes of $\delta$'s we associate the following trees:
\begin{eqnarray}
\delta^k_{ji} &=& \; \bullet\;^k_{ji}~, \nonumber
\\
\delta^k_{ji,l} &=& \parbox{8mm}{\begin{picture}(6,11)
\put(1,8){$\bullet\;^k_{ji}$}
\put(2,9){\line(0,-1){7}}
\put(1,1){$\bullet\;_l$}
\end{picture}}~,  \nonumber
\\
\delta^k_{ji,lm} &=& \parbox{8mm}{\begin{picture}(6,18)
\put(1,15){$\bullet\;^k_{ji}$}
\put(2,16){\line(0,-1){14}}
\put(1,8){$\bullet\;_l$}
\put(1,1){$\bullet\;_m$}
\end{picture}} + 
\parbox{14mm}{\begin{picture}(14,11)
\put(4.5,8){$\bullet~^k_{ji}$}
\put(5.5,9){\line(-1,-2){3.5}}
\put(5.5,9){\line(1,-2){3.5}}
\put(1,1){$\bullet\;_l$}
\put(8,1){$\bullet\;_m$}
\end{picture}} ~. \label{tree}
\end{eqnarray}
The rule is obvious. A symbol $\delta^k_{ji,A\ell}$, for $A$ a string
of $|A|$ indices, is obtained from $\delta^k_{ji,A}=\sum_{a=1}^{|A|!}
t_a^{|A|}$ by attaching to each of its trees $t_a^{|A|}$ a new vertex
with label $\ell$ successively to the right of each vertex. The root
(with three indices) remains the same and order is important.

Coproduct and antipode require the definition of cuts of a tree. An
elementary cut along a chosen edge splits a tree into two -- the trees
above (trunk) and below (cut branch) the cut. It is clear that we have
to add 2 indices to complete the root of the cut branch. This will be
a pair of summation indices. We define the action of a cut as the
shift of one index of the vertex above the cut to the first
position of the new root of the cut branch. The remaining position to
complete the root of the cut branch is filled with a summation index
and the same summation index is put into the vacant position of the
trunk. In the case of cutting immediately below the root, we have to
sum over the three possibilities of picking up indices of the root,
adding a minus sign if we pick up the unique upper index. The
following examples illustrate the definition of a cut, where we write
the trunk as the rhs of the tensor product and the cut branch as the
lhs:
\begin{eqnarray}
\parbox{8mm}{\begin{picture}(6,11)
\put(1,8){$\bullet\;^k_{ji}$}
\put(2,9){\line(0,-1){7}}
\put(0,4){---}
\put(1,1){$\bullet\;_l$}
\end{picture}} 
&=& \;\bullet\;^a_{jl}\;\otimes\;\bullet\;^k_{ai}\; 
+ \;\bullet\;^a_{il} \;\otimes\; \bullet\;^k_{ja}\;
- \;\bullet\;^k_{al} \;\otimes\; \bullet\;^a_{ij}~, \nonumber
\\
\parbox{14mm}{\begin{picture}(14,11)
\put(4.5,8){$\bullet~^k_{ji}$}
\put(5.5,9){\line(-1,-2){3.5}}
\put(5.5,9){\line(1,-2){3.5}}
\put(5.5,4){---}
\put(1,1){$\bullet\;_l$}
\put(8,1){$\bullet\;_m$}
\end{picture}} &=& \bullet\;^a_{jm}\; \otimes \;\bullet\;^k_{ai,l}\;+ 
\;\bullet\;^a_{im}\; \otimes \;\bullet\;^k_{ja,l} \;
- \;\bullet\;^k_{am}\; \otimes \;\bullet\;^a_{ji,l}~, \nonumber
\\
\parbox{8mm}{\begin{picture}(6,18)
\put(1,15){$\bullet\;^k_{ji}$}
\put(2,16){\line(0,-1){14}}
\put(0,4){---}
\put(1,8){$\bullet\;_l$}
\put(1,1){$\bullet\;_m$}
\end{picture}} &=&  \;\bullet\;^a_{lm} \;\otimes 
\parbox{8mm}{\begin{picture}(6,11)
\put(1,8){$\bullet~^k_{ji}$}
\put(2,9){\line(0,-1){7}}
\put(1,1){$\bullet~_a$}
\end{picture}}  ~.
\end{eqnarray}
A multiple cut consists of several elementary cuts, where \emph{the
order of cuts is from top to bottom and from left to right}. An
\emph{admissible} cut is a multiple cut such that on the path from any
vertex to the root there is at most one elementary cut. The product of
all cut branches forms the lhs of the tensor product, whereas the
trunk alone containing the old root serves as the rhs.

The purpose of these definitions is to give an explicit formula for
coproduct and antipode. Indeed, by induction one can prove the
following:
\begin{Proposition}
The coproduct of $\delta^k_{ji,A} = \sum_{a=1}^{|A|!} t^{|A|}_a$ is
given by
\begin{equation}
\Delta (\delta^k_{ji,A})= \delta^k_{ji,A} \otimes 1+1 \otimes
\delta^k_{ji,A} +  \sum_{a=1}^{|A|!} \sum_{\mathcal{C}} P^{\cal
C}(t^{|A|}_a) \otimes R^\mathcal{C}(t^{|A|}_a)~,
\label{PR}
\end{equation}
where for each $t^{|A|}_a$ the sum is over all admissible cuts
$\mathcal{C}$ of $t^{|A|}_a$. In eq.\ (\ref{PR}),
$R^\mathcal{C}(t^{|A|}_a)$ is the trunk and $P^\mathcal{C}(t^{|A|}_a)$
the product of cut branches obtained by cutting $t^{|A|}_a$ via the
multiple cut $\mathcal{C}$.
\end{Proposition}
\textit{Proof.} We start from 
\begin{eqnarray*}
\Delta (\delta^k_{ji,A\ell}) &=&
[\Delta(\delta^k_{ji,A}),\Delta(X_\ell)] 
= \delta^k_{ji,A\ell} \otimes 1 + 1 \otimes \delta^k_{ji,A\ell} +
R^k_{ji,A\ell}~, 
\\
R^k_{ji,A\ell} &=& [X_\ell \otimes 1 + 1 \otimes X_\ell  , R^k_{ji,A}] 
+ [\delta^m_{n\ell} \otimes Y^n_m , R^k_{ji,A} + (1 \otimes 
\delta^k_{ji,A})] \in \mathcal{H} \otimes \mathcal{H}~.
\end{eqnarray*}
By definition of the tree, the commutator with $X_\ell$ attaches a
vertex $\ell$ successively to all previous vertices, where $X_\ell
\otimes 1$ attaches to the cut branches and $1 \otimes X_\ell$ attaches
to the trunk. Next, the commutator with $\delta^m_{n\ell} \otimes
Y^n_m$ puts for each vertex of the trunk (due to the commutator with
$Y$) a cut branch consisting of a single vertex to the lhs of the
tensor product. Both contributions together yield precisely all
admissible cuts of the trees corresponding to $\delta^k_{ji,A\ell}$.
\qed
\vskip 1ex

The antipode is obtained by applying the antipode axiom $m \circ (S
\otimes {\rm id}) \circ \Delta =0$ to (\ref{PR}). By recursion one
proves
\begin{Proposition}
The antipode $S$ of $\delta^k_{ji,A} =\sum_{a=1}^{|A|!}
t^{|A|}_a$ is given by 
\begin{equation}
S(\delta^k_{ji,A}) = - \delta^k_{ji,A} - \sum_{a=1}^{|A|!} 
\sum_{\mathcal{C}_a}  (-1)^{|\mathcal{C}_a|}\; 
P^{\mathcal{C}_a}(t^{|A|}_a)\,R^{\mathcal{C}_a}(t^{|A|}_a)~,
\label{Sprp}
\end{equation}
where the sum is over the set of all non-empty multiple cuts ${\cal
C}_a$ of $t^{|A|}_a$ (multiple cuts on paths from bottom to the root
are allowed) consisting of $|\mathcal{C}_a|$ individual cuts. \qed
\end{Proposition}

\section{Feynman graphs and rooted trees}

In a perturbative quantum field theory it is convenient to symbolize
contributions to Green's functions by Feynman graphs. These Feynman
graphs stand for analytic expressions of momentum variables. Internal
momentum variables have to be integrated out. Very often some of these
integrations formally yield infinity. The art of obtaining meaningful
results out of these integrals is called renormalization. A central
problem is the existence of subdivergences which cannot be regularized
by a simple subtraction of the divergent part. Bogoliubov \cite{bs}
found a recursion formula for the regularization of Feynman graphs
with subdivergences and Zimmermann gave an explicit solution -- the
forest formula \cite{z}.

In 1997 Dirk Kreimer discovered \cite{k1} that there is the structure
of a Hopf algebra behind this art of renormalization, with the
combinatorics of the forest formula produced by the
antipode. Kreimer's idea was to visualize the divergence structure of
Feynman graphs in terms of parenthesized words, which are in 1:1
correspondence to rooted trees \cite{ck1}. Let us exemplify this idea
by a Feynman graph from QED:
\begin{equation}
\parbox{70mm}{\epsfig{file=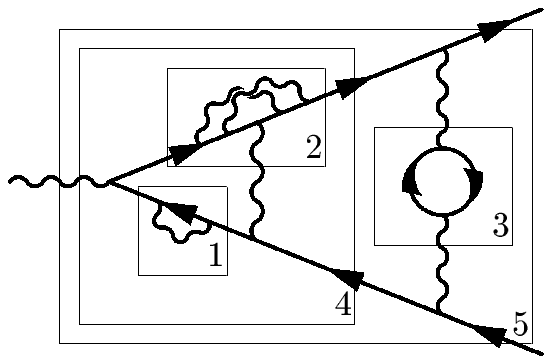}} = \quad 
\parbox{25mm}{ 
\begin{picture}(20,33)
\put(14,19){$\bullet$}
\put(17,20){$v_5$}
\put(15,20){\line(-1,-1){5}}
\put(9,14){$\bullet$}
\put(5,16){$v_4$}
\put(15,20){\line(1,-1){5}}  
\put(19,14){$\bullet$}
\put(22,14){$p_3$}
\put(10,15){\line(-1,-1){5}}  
\put(4,9){$\bullet$}
\put(0,9){$s_1$}
\put(10,15){\line(1,-1){5}}
\put(14,9){$\bullet$}
\put(17,9){$v_2$}
\end{picture}}
\end{equation}
Straight lines stand for fermions and wavy lines for bosons, and the
boxes contain divergent sectors. A criterion for superficial
divergence of a region confined in a box is power counting. If a box
has $n_B$ bosonic and $n_F$ fermionic outgoing legs, the power
counting degree of divergence $d$ is (in four dimensions) defined by
$d:=4-n_B - \frac{3}{2} n_F \geq 0$.  Owing to symmetries the actual
degree of divergence of one graph or a sum of graphs can be lower than
$d$, see \cite{iz}. The construction of the rooted tree from the
Feynman graphs with identified divergent sectors is clear: The
outermost (superficial) divergence $(5)$ is the root $v_5$. The box
$(5)$ contains the boxes $(3)$ and $(4)$ as immediate subdivergences,
hence we connect two vertices $p_3$ and $v_4$ directly to the root
$v_5$. The box $(4)$ contains the subdivergences $(1)$ and $(2)$, so
we attach the vertices $s_1$ and $v_2$ to $v_4$. This works as long as
there are no overlapping divergences, which must be resolved before in
terms of disjoint and nested ones and lead to a sum of rooted trees
\cite{kw,k3}.

Having identified the trees to Feynman graphs, it are the same cutting
operations on trees as before which give us coproduct and
antipode. Here, a cut splits a Feynman graph into several
subgraphs -- a standard operation in renormalization. It is very
remarkable that the antipode obtained in this way reproduces the
combinatorics of renormalization \cite{k1}. These surprising facts
have been extended to a complete renormalization of a toy model
\cite{k2}, which we review in the next section.

Before, let us ask an interesting question: What is the role of the
operators $\delta^k_{ji,\ell_1\dots\ell_n}$ in quantum field theory,
and what is the meaning of the individual trees for diffeomorphisms? I
am not aware of an answer, but there is an interesting observation
\cite{rw} concerning the relation of the \emph{decorated} rooted trees
(\ref{tree}) to Feynman graphs. The trees emerging from the
Connes--Moscovici Hopf algebra are decorated by spacetime indices
(three for the root) whereas in QFT the decoration is a label for
divergent Feynman graphs without subdivergences.  Although the
operators $\delta$ are invariant under permutation of the indices
after the comma, for instance $\delta^k_{ji,lm}=\delta^k_{ji,ml}$, see
(\ref{flat}), this symmetry is lost on the level of individual
trees. That leads us to speculate that \textit{the sum of Feynman graphs
according to the collection of rooted trees to $\delta$'s has more
symmetry than the individual Feynman graphs.}  This should be checked
in QFT calculations.  Another interpretation would be the observation
from (\ref{tree})
\begin{equation}
\parbox{8mm}{\begin{picture}(6,18)
\put(1,15){$\bullet~^k_{ji}$}
\put(2,16){\line(0,-1){14}}
\put(1,8){$\bullet~_l$}
\put(1,1){$\bullet~_m$}
\end{picture}} + 
\parbox{14mm}{\begin{picture}(14,11)
\put(4.5,8){$\bullet~^k_{ji}$}
\put(5.5,9){\line(-1,-2){3.5}}
\put(5.5,9){\line(1,-2){3.5}}
\put(1,1){$\bullet~_l$}
\put(8,1){$\bullet~_m$}
\end{picture}} - 
 \parbox{8mm}{\begin{picture}(6,18)
\put(1,15){$\bullet~^k_{ji}$}
\put(2,16){\line(0,-1){14}}
\put(1,8){$\bullet~_m$}
\put(1,1){$\bullet~_l$}
\end{picture}} -
\parbox{14mm}{\begin{picture}(14,11)
\put(4.5,8){$\bullet~^k_{ji}$}
\put(5.5,9){\line(-1,-2){3.5}}
\put(5.5,9){\line(1,-2){3.5}}
\put(1,1){$\bullet~_m$}
\put(8,1){$\bullet~_l$}
\end{picture}} =0 ~,
\label{rel}
\end{equation}
which could possibly be regarded as a relation between Feynman graphs
similar to those derived in \cite{k4}. According to a private
communication by Kreimer, (\ref{rel}) is satisfied in QFT for the
leading divergences, as it can be derived from sec.\ V.C in
\cite{kd}. For non-leading singularities there will be (probably
systematic) modifications.

In mathematics, Connes and Kreimer extended the investigation of the
commutative Hopf subalgebra $\mathcal{H}^1$ in \cite{cm2} to the level
of individual trees \cite{ck1}. They showed that the Hopf algebra of
rooted trees $\mathcal{H}_R$ is the solution of a universal problem in
Hochschild cohomology.  We recall \cite{cm2} that $\mathcal{H}^1$ is
the dual of the enveloping algebra of the Lie algebra $\mathcal{L}^1$
of formal vector fields on ${\bb R}$ vanishing to order 2 at the
origin, and that $\mathcal{H}^1$ itself is isomorphic to the Hopf
algebra of coordinates on the group of diffeomorphisms of ${\bb R}$ of
the form $\psi(x)=x+o(x)$. By analogy, Connes and Kreimer regard
${\cal H}_R$ as the Hopf algebra of coordinates on a nilpotent formal
group $\mathcal{G}$ whose Lie algebra $\mathcal{L}\sp 1$ they succeed
to compute. This group was recently found to be related to the Butcher
group in numerical analysis \cite{b}. It will certainly contain
precious information for quantum field theory because the antipode in
$\mathcal{H}_R$ governing renormalization is the dual of the inversion
operation in $\mathcal{G}$. Renormalization seems to provide a new
mathematical calculus which generalizes differential calculi.

\section{A toy model: iterated integrals}

In the spirit of Kreimer \cite{k2} we are going to give the reader a
feeling for renormalization by considering a toy model. The toy model
is given by iterated divergent integrals, in close analogy to QFT. The
only difference is that the integrals are very simple to compute.

Let us take the  integral 
\begin{equation}
\Gamma^1(t) = \int_t^\infty \frac{dp_1}{p_1^{1+\epsilon}}~,
\end{equation}
which diverges logarithmically for $\epsilon \to 0$. We can regard it
as the analytic expression to the Feynman graph
\[
\bullet ~= \quad \parbox{18mm}{\begin{picture}(18,12)
\put(0,6){\line(3,1){17}}
\put(0,6){\line(3,-1){17}}
\put(0,6){\line(-1,0){4}}
\put(12,10){\line(0,-1){8}}
\end{picture}}
\]
To a Feynman graph with subdivergence there corresponds an iterated
integral:
\begin{eqnarray}
\parbox{8mm}{\begin{picture}(6,11)
\put(1,8){$\bullet$}
\put(2,9){\line(0,-1){7}}
\put(1,1){$\bullet$}
\end{picture}} ~= \quad
\parbox{18mm}{\begin{picture}(18,12)
\put(0,6){\line(3,1){18}}
\put(0,6){\line(3,-1){18}}
\put(0,6){\line(-1,0){4}}
\put(15,11){\line(0,-1){10}}
\put(12,10){\line(0,-1){8}}
\end{picture}}
\quad
&\longleftrightarrow&
\quad
\Gamma^2(t) = \int_t^\infty \frac{dp_1}{p_1^{1+\epsilon}} 
\int_{p_1}^\infty \frac{dp_2}{p_2^{1+\epsilon}} ~, \nonumber
\\*
\parbox{8mm}{\begin{picture}(6,18)
\put(1,15){$\bullet$}
\put(2,16){\line(0,-1){14}}
\put(1,8){$\bullet$}
\put(1,1){$\bullet$}
\end{picture}} ~= \quad
\parbox{18mm}{\begin{picture}(18,12)
\put(0,6){\line(3,1){18}}
\put(0,6){\line(3,-1){18}}
\put(0,6){\line(-1,0){4}}
\put(15,11){\line(0,-1){10}}
\put(12,10){\line(0,-1){8}}
\put(9,9){\line(0,-1){6}}
\end{picture}}
\quad
&\longleftrightarrow&
\quad
\Gamma^3(t) = \int_t^\infty \frac{dp_1}{p_1^{1+\epsilon}} 
\int_{p_1}^\infty \frac{dp_2}{p_2^{1+\epsilon}} 
\int_{p_2}^\infty \frac{dp_3}{p_3^{1+\epsilon}} ~.
\end{eqnarray}
Clearly, these iterated integrals form a Hopf algebra of rooted trees
without side branches, and the coproduct is given by the admissible
cuts of the trees. The renormalization of these integrals requires an
algebra homomorphisms $\phi_a$ on iterated integrals, which represents
a certain way of evaluation under ``a set of conditions $a$''. For our
purpose we take
\begin{equation}
\phi_a\Big(\prod_{i \in I} \Gamma^i(t)\Big) := 
\prod_{i \in I} \Gamma^i(a)~,
\end{equation}
the evaluation of the integrals at $t=a$. In QFT, $a$ should be
regarded as an energy scale, and $\phi_a$ evaluates the Feynman graphs
at this scale. 

The essential idea \cite{k2} is now to consider the \emph{convolution
product} of these homomorphisms, defined via the Hopf algebra
structure:
\begin{equation}
(\phi \star \psi)(h) := m \circ (\phi \otimes \psi) \circ \Delta (h)~,
\qquad h \in \mathcal{H}~.
\end{equation}
The antipode axiom can be written in the compact form $S \star
\mathrm{id} = 1\,\varepsilon$. It is however more interesting to
consider the following modification:
\begin{equation}
\varepsilon_{a,b} = S_a \star \mathrm{id}_b := (\phi_a \circ S) \star 
\phi_b~.
\end{equation}
Due to the Hopf algebra properties, the $\varepsilon_{a,b}$ satisfy a
groupoid law. We give the derivation in full detail, using
1) associativity of $m$ and coassociativity of
$\Delta$, 2) the antipode axiom, 3) homomorphism property of $\phi$,
4) $\phi \circ 1 \epsilon = 1\epsilon$, 5) the counit axiom:
\begin{eqnarray*}
\varepsilon_{a,b} \star \varepsilon_{b,c} 
&&= m \circ \Big( 
\big( m \circ \big(S_a \otimes \phi_b) \circ \Delta \big) \big)
\otimes \big(
m \circ \big( S_b \otimes \phi_c) \circ \Delta \big) \big)\Big) 
\circ \Delta
\\
&&=  m \circ (m \otimes m) \circ \Big( 
S_a \otimes \phi_b \otimes S_b \otimes
\phi_c\Big) \circ (\Delta \otimes \Delta) \circ \Delta
\\
&&= m \circ (\mathrm{id} \otimes m ) \circ (m \otimes \mathrm{id}
\otimes \mathrm{id}) \circ \Big( 
S_a \otimes \phi_b \otimes S_b \otimes
\phi_c\Big) \circ \\
&& \hspace*{6.5cm} \circ (\Delta \otimes \mathrm{id} \otimes \mathrm{id}) 
\circ (\mathrm{id} \otimes \Delta) \circ \Delta
\\
&&=^1 m \circ (m \otimes \mathrm{id}) \circ (m \otimes \mathrm{id}
\otimes \mathrm{id}) \circ \Big( 
S_a \otimes \phi_b \otimes S_b \otimes
\phi_c\Big) \circ \\
&& \hspace*{6.5cm} \circ (\Delta \otimes \mathrm{id} \otimes \mathrm{id}) 
\circ (\Delta \otimes \mathrm{id}) \circ \Delta
\\
&&= m \circ (\Big( m \circ (m \otimes \mathrm{id}) \circ 
( S_a \otimes \phi_b \otimes S_b) \circ (\Delta \otimes \mathrm{id}) 
\circ \Delta \Big) \otimes \phi_c ) \circ \Delta
\\
&&=^1 m \circ (\Big( m \circ (\mathrm{id} \otimes m) \circ 
( S_a \otimes \phi_b \otimes S_b) \circ (\mathrm{id} \otimes \Delta) 
\circ \Delta \Big) \otimes \phi_c ) \circ \Delta
\\
&&= m \circ (\Big( m \circ \{S_a \otimes \Big(m \circ 
(\phi_b \otimes \phi_b) \circ (\mathrm{id} \otimes S) \circ \Delta\Big) 
\} \circ \Delta \Big) \otimes \phi_c ) \circ \Delta
\\
&&=^{2,3} m \circ (\Big( m \circ \{S_a \otimes \Big(
\phi_b \circ 1\epsilon \Big) \} 
\circ \Delta \Big) \otimes \phi_c ) \circ \Delta
\\
&&=^4 m \circ (\Big( m \circ (S_a \otimes \mathrm{id}) \circ (\mathrm{id} 
\otimes 1 \epsilon ) \circ \Delta \Big) 
\otimes \phi_c ) \circ \Delta
\\
&&= m \circ (m \otimes \mathrm{id}) \circ (S_a \otimes \mathrm{id}
\otimes \phi_c) \circ (\mathrm{id} 
\otimes 1\epsilon \otimes \mathrm{id}) \circ (\Delta \otimes \mathrm{id})
\circ \Delta
\\
&&=^{1,4} m \circ ( \mathrm{id} \otimes m) \circ (S_a \otimes \phi_c 
\otimes \phi_c) \circ (\mathrm{id} 
\otimes 1\epsilon \otimes \mathrm{id}) \circ (\mathrm{id} \otimes \Delta)
\circ \Delta
\\
&&=^3 m \circ (S_a \otimes \phi_c ) \circ (\mathrm{id} \otimes 
\big( m \circ (1\epsilon \otimes \mathrm{id}) \circ \Delta \big)) \circ
\Delta  
\\
&&=^5 m \circ (S_a \otimes \phi_c ) \circ \Delta
\\
&&= \varepsilon_{a,c}~.
\end{eqnarray*}

We apply now the $\varepsilon_{a,b}$ operation to the divergent
integrals to compute $\varepsilon_{a,b}(\Gamma^i(t))=\Gamma^i_{a,b}$:
\begin{eqnarray*}
\Gamma^1_{a,b} &=& m \circ (\phi_a \otimes \phi_b) \circ (S \otimes
\mathrm{id} ) \circ \Delta (\bullet)
\\*
&=& m \circ (\phi_a \otimes \phi_b) 
\circ \big( - \bullet \otimes 1 + 1 \otimes \bullet )
\\*
&=& - \Gamma^1(a) + \Gamma^1(b) = \displaystyle \int_b^a
\frac{dp}{p^{1+\epsilon}} ~.
\end{eqnarray*}
The result $\Gamma^1_{a,b}$ is finite for $\epsilon \to 0$ and
vanishes for $a=b$. We proceed with the next integral, using the
definition of $\Delta$ as given by the admissible cuts and $S$ as
given by all cuts (with sign from the number of elementary cuts) of
the graphs:
\begin{eqnarray*}
\Gamma^2_{a,b} &=& m \circ (\phi_a \otimes \phi_b) \circ (S \otimes
\mathrm{id} ) \circ \Delta \Big(
\parbox{4mm}{\begin{picture}(4,11)
\put(1,8){$\bullet$}
\put(2,9){\line(0,-1){7}}
\put(1,1){$\bullet$}
\end{picture}} \Big)
\\*
&=& m \circ (\phi_a \otimes \phi_b) \circ \Big( 
S\Big(\parbox{4mm}{\begin{picture}(4,11)
\put(1,8){$\bullet$}
\put(2,9){\line(0,-1){7}}
\put(1,1){$\bullet$}
\end{picture}} \Big) \otimes 1 + S(\bullet) \otimes \bullet + 1
\otimes \parbox{4mm}{\begin{picture}(4,11)
\put(1,8){$\bullet$}
\put(2,9){\line(0,-1){7}}
\put(1,1){$\bullet$}
\end{picture}} \Big)
\\
&=& m \circ (\phi_a \otimes \phi_b) \circ \Big( 
- \parbox{4mm}{\begin{picture}(4,11)
\put(1,8){$\bullet$}
\put(2,9){\line(0,-1){7}}
\put(1,1){$\bullet$}
\end{picture}} \otimes 1
+ \bullet \bullet \otimes 1 - \bullet \otimes \bullet 
+ 1 \otimes \parbox{4mm}{\begin{picture}(4,11)
\put(1,8){$\bullet$}
\put(2,9){\line(0,-1){7}}
\put(1,1){$\bullet$}
\end{picture}} \Big) 
\\
&=& - \Gamma^2(a) + \Gamma^1(a) \Gamma^1(a) - \Gamma^1(b) \Gamma^1(a) +
\Gamma^2(b) 
\\
&=& \displaystyle \Big( -\int_a^\infty \int_{p_1}^\infty + 
\int_a^\infty \int_a^\infty - \int_b^\infty \int_a^\infty 
+ \int_b^\infty \int_{p_1}^\infty \Big) 
\frac{dp_1}{p_1^{1+\epsilon}} 
\frac{dp_2}{p_2^{1+\epsilon}} 
\\*
&=&  \displaystyle \int_b^a \frac{dp_1}{p_1^{1+\epsilon}} \int_{p_1}^a 
\frac{dp_2}{p_2^{1+\epsilon}}~.
\end{eqnarray*}
Again, the result is finite. Note that in $\bullet \otimes \bullet$
the root which stands for the $p_1$ integration is the right vertex
and hence is evaluated at $t=b$. The computation for $\Gamma^3_{a,b}$
is left as an exercise. 

From the identity $\varepsilon_{a,b} \star \varepsilon_{b,c} = 
\varepsilon_{a,c}$ and the coproduct rule given by admissible cuts of
a tree without side branches we get Chen's Lemma \cite{chen}:
\begin{equation}
\Gamma^i_{a,c} = \Gamma^i_{a,b} + \Gamma^i_{b,c} + \sum_{j=1}^{i-1} 
\Gamma^j_{a,b} \Gamma^{i-j}_{b,c}~.
\end{equation}
For $i=2$ it reads
\[
\int_c^a \frac{dp_1}{p_1} \int_{p_1}^a \frac{dp_2}{p_2} = 
\int_b^a \frac{dp_1}{p_1} \int_{p_1}^a \frac{dp_2}{p_2} +
\int_c^b \frac{dp_1}{p_1} \int_{p_1}^b \frac{dp_2}{p_2} +
\int_c^b \frac{dp_1}{p_1} \int_b^a \frac{dp_2}{p_2} ~.
\]

The purpose of these considerations was the renormalization of a
QFT. Let us assume a theory where all contributions to the coupling
constant come from the following ladder diagrams:
\[
\begin{array}{ccccccccccc}
\parbox{14mm}{\begin{picture}(14,12)
\put(4,6){\line(2,1){10}}
\put(4,6){\line(2,-1){10}}
\put(4,6){\line(-1,0){4}}
\put(3,5){$\bullet$}
\end{picture}}
&=&
\parbox{14mm}{\begin{picture}(14,12)
\put(4,6){\line(2,1){10}}
\put(4,6){\line(2,-1){10}}
\put(4,6){\line(-1,0){4}}
\end{picture}}
&+& 
\parbox{22mm}{\begin{picture}(22,12)
\put(4,6){\line(3,1){18}}
\put(4,6){\line(3,-1){18}}
\put(4,6){\line(-1,0){4}}
\put(16,10){\line(0,-1){8}}
\end{picture}}
&+& 
\parbox{22mm}{\begin{picture}(22,12)
\put(4,6){\line(3,1){18}}
\put(4,6){\line(3,-1){18}}
\put(4,6){\line(-1,0){4}}
\put(19,11){\line(0,-1){10}}
\put(16,10){\line(0,-1){8}}
\end{picture}}
&+& 
\parbox{22mm}{\begin{picture}(22,12)
\put(4,6){\line(3,1){18}}
\put(4,6){\line(3,-1){18}}
\put(4,6){\line(-1,0){4}}
\put(19,11){\line(0,-1){10}}
\put(16,10){\line(0,-1){8}}
\put(13,9){\line(0,-1){6}}
\end{picture}}
&+& \dots
\\
\Gamma &=& \Gamma^0 &+& \Gamma^1 &+& \Gamma^2 &+& \Gamma^3 &+& \dots
\end{array}
\]
Formally, this series evaluates to infinity, but this infinity can be
renormalized to a finite but \emph{undetermined} value. That value has
to be adapted to experiment and yields a \emph{normalization
condition}. At some energy scale $a$ we are allowed to fix the
coupling constant $\Gamma_a = \Gamma^0(a)$. But suppose we measure now
the value of the coupling constant at another energy scale $b$. The
normalization condition is fixed so that in the
diagrams we have to use in all vertices the normalized coupling
constant $\parbox{7mm}{\begin{picture}(7,8) \put(3,4){\line(2,1){4}}
\put(3,4){\line(2,-1){4}} \put(3,4){\line(-1,0){3}}
\end{picture}} = \Gamma_a$. Since the renormalization removing the
infinities was scale dependent, the loop diagrams $\Gamma^i$ now give
a contribution, and this contribution is precisely $\Gamma^i_{a,b}$.
Hence,
\begin{equation}
\Gamma_b = \Gamma_a + \Gamma^1_{a,b} + \Gamma^2_{a,b} + \Gamma^3_{a,b}
+ \dots 
\label{ren}
\end{equation}
Assuming the series converges, we get a finite shift of the coupling
constant. In realistic quantum field theories, the agreement of this
value with experiment is overwhelming. In particular, in first order
we recover the familiar logarithmic energy dependence of the coupling
constant. We also learn from (\ref{ren}) that one can completely avoid
talking about infinities. 

As it is clear from our model, the running coupling constants
resulting from renormalization are governed by the Hopf algebra
structure together with the convolution product. The Hopf algebra
structure not only produces the combinatorics of the forest formula,
it also allows to compare different renormalization schemes, which
arise from each other by a finite re-normalization. The theory is
consistent without a preferred scale or preferred renormalization
scheme. They are always related by the convolution identity
$\varepsilon_{a,c} = \varepsilon_{ab} \star \varepsilon_{bc}$, where
$a,b,c$ stand for parameterizations of different renormalization
schemes. Applications of these ideas to QFT calculations are starting
\cite{kd}. 

\section*{Acknowledgements}

I am grateful to the organizers of the Hesselberg'99 conference,
Florian Scheck, Harald Upmeier and Wend Werner, for the invitation and
the possibility to present these ideas.  It is a pleasure to thank my
colleagues Bruno Iochum, Thomas Krajewski, Serge Lazzarini, Thomas
Sch\"ucker and Daniel Testard for collaboration and numerous
discussions. Finally, I would like to thank Alain Connes and Dirk
Kreimer for important advice at various stages of my study of the
Hopf algebras.

\end{document}